# The Effect of Native Language on Internet Usage

Neil Gandal (Tel Aviv University)

Carl Shapiro (University of California at Berkeley)

September 2001

## ABSTRACT

In this paper, we empirically explore the relationship between native language and use of the Internet. The ultimate economic and social questions we explore are: (1) how native language affects use of the Internet, both in total and by type of Web site; (2) whether English is likely to retain its "first-mover advantage" on the Web in terms of the language employed by Web sites; and (3) whether the Internet ultimately will accelerate the movement to English as a global language.

Our goal is to distinguish between the following two hypotheses: (A) The Internet will remain disproportionately in English and will, over time, cause more people to learn English as second language and thus solidify the role of English as a global language. This outcome will prevail even though there are more native Chinese and Spanish speakers than there are native English speakers. (B) As the Internet matures, it will more accurately reflect the native languages spoken around the world (perhaps weighted by purchasing power) and will not promote English as a global language.

English's "early lead" on the web is more likely to persist if those who are *not* native English speakers frequently access the large number of English language web sites that are currently available. In that case, many existing web sites will have little incentive to develop non-English versions of their sites, and new sites will tend to gravitate towards English. The key empirical question, therefore, is whether individuals whose native language is *not* English use the Web, or certain types of Web sites, less than do native English speakers. In order to examine this issue empirically, we employ a unique data set on Internet use at the individual level in Canada from Media Metrix. Canada provides an ideal setting to examine this issue because English is one of the two official languages.

Our preliminary results suggest that English web sites are *not* a barrier to Internet use for French-speaking Quebecois. These results are consistent with the scenario in which the Internet will promote English as a global language.

# 1. Introduction: Network Effects, First-Mover Advantages, and Language

In recent years, English has become the *de facto* standard for business and academic communication and has to some degree attained the status of a global language. English is the official language of the Asian trade group ASEAN and the official language of the European Central Bank, despite the fact that the bank is in (Frankfort) Germany and neither the U.K. nor Ireland are members of the European Monetary Union.[1] Several public schools in Zurich, Switzerland are now teaching some of the elementary school subjects in English. This is occurring in a country where there are four official languages --French, German, Italian, and Romansch. In a recent European Union survey, 70 percent agreed with the notion that everybody should speak English.[2]

In this paper , we examine how native language affects Internet use. The goal is to determine whether the Internet is likely to remain disproportionately English and thus whether the Internet will accelerate the movement to English as a global language.

Currently there is much more Internet content available in English than in other languages. A recent estimate by Global Reach indicates that nearly 70 percent of all Internet content is currently in English. Japanese and German follow with approximately 6 percent each.[3] The Internet certainly is an effective instrument for circulating English around the world.

On the other hand, it is quite possible that several languages will have a large critical mass of Internet content, so that English's role as a global language will diminish: Although 44 percent of current Web users (March 2001) are native English speakers,[4] web use is currently growing faster among non-native English speakers. Indeed, it is estimated that by 2003, only 29 percent of all Internet users will be native English speakers. Additionally because of low transaction costs, the Internet is ideal for bringing together members of small groups like speakers of Frisian, which is spoken by approximately 500,000 people throughout the world.

---

[1] Wallraff, B., What Global Language, *Atlantic Monthly*, November 2000, p.53-66.

[2] Daley, S., "In Europe, Some Fear National Languages are Endangered," NY Times, 4/16/01.

[3] See http://www.euromktg.com/globstats/refs.php3.

[4] Source: See Global Reach at http://www.glreach.com/globstats/. Following English, 9.0 percent of all Internet users are native Chinese speakers, while 8.6 percent are Japanese, and 6.1 percent are German.

Concern that national languages are becoming endangered might lead policy makers to require websites to be in the domestic national language. France already has many laws in place that protect the French language. Quebec requires all websites in that province be available in French. In Brazil, which has the largest Internet industry in South America, a bill was recently introduced that would prohibit the introduction and use of foreign words.[5]

Additionally, another standards war is already brewing on Internet domain names, which until quite recently exclusively used romance language alphabets. Currently two groups, Verisign Inc. and China's Internet Authority now issue Chinese language domain names using two incompatible systems. This means that in order two operate in both cyber-spaces, businesses would have to register with both authorities.[6] Pindar Wong, the former vice president of the Internet Corporation for Assigned Names and Numbers (ICANN), the global body that governs the Internet indicated that the two incompatible systems risk "Balkanization of the Internet, dividing the Internet up into islands of connectivity."[7]

A key determinant of whether the Internet will move towards Balkanization (critical mass of content in many languages) or standardization is whether the "first-mover advantage" (of significant web content in English) will encourage non-English speakers to use English language web sites. In such a case, existing content providers would have little incentive to offer non-English versions of their sites, and new sites would have a strong incentive to provide their content in English. Such a first mover advantage may lead to a "bandwagon" because there are network effects in language: learning a second language is more valuable, the more widely that language is used.

A network effect exists when the value that consumers place on a particular product increases as the total number of consumers who use identical or compatible goods increases. In the case of an actual (or physical) network, such as the telephone or email network, the value of the network depends on the total number of subscribers who have access to the network. Since languages are in part communication technologies, the value of a language network increases in the number of speakers and users of that language. Languages are perfect substitutes, but they are incompatible in the sense that two individuals can talk with each other only if they both speak the same language.[8] Languages, as a type of communications networks, clearly are subject to strong direct network effects.

---

[5] NY Times, 5/14/01, "English is Spoken Here…Too Much, Some Say," by Larry Rohter.
[6] Wall Street Journal, 11/30/2000, "Will Language Wars Balkanize the Web", by Gren Manuel and Leslie Chang.

In the case of virtual networks, that are not linked physically, a network effect arises from positive feedback from complementary goods.[9] The positive feedback mechanism works as follows: the value of the base product (such as a DVD players) is enhanced as the variety of (compatible) complementary products (content available on DVD disc) increases; hence consumers will be more likely to purchase a base product with many compatible complementary products. The variety of complementary products, in turn, will depend on the total number of consumers that purchase the base product. As the number of consumers that purchase the base product increases, there is a greater demand for compatible complementary products. This increases the profitability of supplying complementary products. Since there are typically fixed or sunk entry costs, production of the complementary products is characterized by increasing returns to scale. So, more complementary products will be produced or developed for a base product with a large share of the market. This further enhances the value of the base product, causing positive feedback in the system: an increase in the sales of the base product leads to more compatible complementary products, which further increases (the value of and) sales of the base product. See Chou and Shy (1990) and Church and Gandal (1992).

Languages are also subject to such indirect or virtual network effects. The value of speaking English increases as the number of English language web sites (or other content, such as books,

---

[7] Ibid.

[8] Of course, when there is no common language, people can still communicate in non verbal ways, such as gestures, expressions, etc.

[9] Examples of virtual networks in which the value of the "base" product increases as the variety of complementary products increases include computer operating systems, videocassette recorders (VCRs), compact disc players (CD-players), and Digital Versatile Disc players (DVD-players).

Language TPRC.doc, 9/5/01 10:31 PM, Page 4 of 4

magazines, or movies) increases. This will lead to an increase in the number of non-English speakers learning English in order to have access to the English language web sites, since individuals who speak English will have more web sites to use. This in turn will lead to an increase in the number of English language web sites.

Markets in which there are network effects are often characterized by tipping: once a system has gained an initial lead, there is a snowball effect. Katz and Shapiro (1994, p. 105) note that positive feedback means that there is a "natural tendency towards de facto standardization." Hence it is possible that the first-mover advantage of English may result in English remaining the dominant language on the web, while other languages will end up serving niche markets. This could occur even though there are more native Chinese and Spanish speakers than there are native English speakers. The outcome will depend, in large part, on the strength of the virtual network effect: Does the large number of English language web sites encourage non English speakers to learn English so that they can access them?[10]

The use of language on the Internet can fruitfully be viewed as a *co-adoption process*.[11] Here "adoption" means use of a particular language; we are thus thinking of *language training and use* as comparable to *technology adoption* decisions that have been extensively studied. The operator of a web site "adopts" a language by offering its site in that language. Likewise, an individual "adopts" a language by learning that language.

More specifically, focusing on the *decisions* made by web sites and users, we can examine the dynamics of language adoption over three time frames. In the *short-term* (day to day), individuals decide – based in part on their language skills and in part on the available offerings in different languages – which web sites to visit, how long to stay at these sites, and whether to engage in commercial transactions. These decisions determine actual Internet usage by different

---

[10] In the case of language, translation is an ex-post substitute for compatibility. If translation utilities worked well, the issue of language would likely be less important. Given the subtleties involved in language, translation by artificial intelligence is in its infancy and currently works quite poorly.

[11] Co-adoption processes are common when virtual network effects are present. For example, when there were two rival incompatible formats for 56k modems, Internet Service Providers and consumers selected modem formats, each influenced by the other group's decisions. This same dynamic arises as well in various server-client

individuals and groups (such as the group of native French speakers). In the *medium term* (over a period of several months to a year or two), operators of web sites decide which language to use for their site, and whether to offer their sites in multiple languages (if permitted this choice by their local governments). These decisions are driven in large part by the amount of traffic that a site expects to attract in one language or another, plus the *incremental traffic* that a site expects to attract by offering its content in multiple languages. Over the *long term* (more than one to two years), individuals (and their parents and teachers) make decisions about which languages to learn. This decisions are driven in part by the desire to access certain content, as well as the desire to communicate directly with others speaking other languages.

In subsequent research, we will to develop a simple theoretical model that captures these three inter-related decisions (visiting web sites, creating content in different languages, and learning languages) that take place continually but over different time frames. We believe that this theoretical treatment, building on the literature on technology adoption and network effects, will support the following general line of reasoning: if in the short term non-native English speakers routinely and extensively use English-language sites, the incentives over the medium term for web sites to make their content available in other languages is reduced, and as a result the incentive over the long term for individuals to learn English as a second language is enhanced, all of which would support the prediction that the Internet will promote English as a global language. On the other hand, if non-native speakers use the Web less, or conduct fewer transactions over the Web than their native-English counterparts (adjusting for other factors such as income and education), web sites will have stronger incentives to offer sites in languages other than English, and English's first-mover advantage on the Web is more likely to dissipate.

In this paper, we explore this issue empirically. To that end, we have obtained a unique data set on Internet use at the individual level in Canada from Media Metrix. Why Canada? Canada provides an ideal setting to examine this issue because English is one of the two official languages. French is, of course, the other official language of Canada.[12] If French speakers are

---

architectures, and influenced the battle between the Netscape Navigator and Microsoft Internet Explorer browsers, as well as the adoption and use of Sun's Java.

[12] Indeed, many of the studies in the economics of language focus on Canada in general, and Quebec in particular. See Grin and Vaillancourt (1997).

likely to use English content sites where there are few substitutes in French, this suggests that the "first mover advantage" is important and that English may indeed remain the global standard on the Internet.

The attractiveness of beginning with a single country (and a single region within a country) is that there is typically greater heterogeneity across countries than within a single country.[13] Hence in this paper, we focus primarily on Canada and Quebec. But the techniques discussed are applicable to other regions of the world.[14]

Our preliminary results suggest that in most categories, native French speakers in Quebec are not less likely than native English speakers to use the Internet. There are some slight differences in Internet use patterns: native French-speaking Quebecois are somewhat more likely than their English counterparts to use government sites, while English speaking Quebecois are somewhat more likely to spend time at search sites.

We also find that there are some differences in the percent of time spent at English language websites between native French and native English speakers. The differences between the groups in this dimension are less significant for the youngest users (age less than 15) and for the next youngest group of users (ages 15 to 24). This is despite the fact that the youngest native French speakers in Quebec are the least likely to have knowledge of English.

While these results are quite preliminary, they suggest that language is much less of a barrier for younger users. These results are consistent with the scenario in which non-native English speakers extensively use English-language sites, the incentives for web sites to make their content available in other languages is reduced, the incentive for individuals to learn English as a second language is enhanced, and that the Internet will promote English as a global language.

---

[13] For example, in some countries, local phone calls are metered, while in other countries (such as the U.S. and Canada), there is a fixed monthly charge for local service. Additionally, Internet access speed might differ widely by country. It can be difficult to find data on and control for these variables.

[14] At a later stage, we envision examining data from many other countries. There is some casual evidence that the most popular online destinations in Mexico originate in the U.S. and do not target Spanish Speakers. See Heft, D. "Who Rules the Internet in Mexico? Why It's America, The Standard, June 12, 2001, at http://www.thestandard.com/article/0,1902,27096,00.html.

Our work complements and make use of the growing literature on the economics of language. Grin and Vaillancourt (1997) provide an overview of the literature; a nice survey is provided by Grin (1996). The major research area within this field is the empirical relationship between earnings and language attributes. Two recent papers are Chiswick and Miller (1999), and Zavodny (2000). Grin (1990) and Church and King (1993) examine rational language choice and public policy toward bilingualism using theoretical models. Rauch (1999) shows that common language facilitates international trade in differentiated products. Freund and Weinhold (2000) find that increased access to the Internet increases trade flows among developed countries. To the best of our knowledge, there is no work on the relationship between Internet use and language, which is the focus of this study.

The data are described in Section 2. In Section 3, we describe the empirical methodology. Our preliminary results and conclusions are provided in Section 4.

## 2. Data

The project employs a unique data set on Internet use at the individual level in Canada, which comes from Media Metrix, the industry leader in the measurement of Internet use. The data include information on demographics of the user such as income, education, family size, province, etc. Additionally, and this is key for the study, the mother tongue of the user – English or French – is known.

The data on Internet use is very detailed. Complete click-stream data are available for the December 2000 period. These data include a separate entry for each URL that is visited, and include the URL domain, as well as the number of active seconds spent at each URL location.[15]

Data were not collected on the language of the web site. Hence, a computer (spider) program was written to check out the language of each URL domain. Although, there were more than 4 million URL full pages, there are "only" approximately 100,000 unique URL domains; an example of a URL domain is http://www.sfgate.com and an example of a URL full page is http://www.sfgate.com/classifieds/rentals/. Of course, many of the URL domains contain some "interior" pages that are in one language and some interior pages that are in other languages. We employed a crude version of this program for the purposes of obtaining preliminary results.[16]

---

[15] Data on total time is available as well. If a user does not enter a key for 60 seconds, the active time count is halted. Hence for less than sixty seconds at a web site, active time is equal to total time. For time spent on a page beyond 60 seconds, active time is less than total time.

In order to obtain preliminary results, we employed the "basic" spider program on the approximately 40,000 unique URLs for Quebec. Hence, by this method, we are able to assign a language to websites.

We also consider it important to categorize the "type" of website accessed, so we can understand in greater detail how different *types* of Internet usage are influenced by language . Media Metrix did this categorization for approximately 2/3 of the observations, but this only accounted for 25% of the websites. Research assistants classified a large portion of the remaining websites in Quebec, so that nearly 90 percent of all observations in Quebec have been categorized. Further research assistance will be needed to classify the remaining unclassified web sites from Quebec (and the rest of Canada). Using the Media Metrix descriptions of the categories, we use the following categories: (1) Retail, Business, Finance; (2) Entertainment, News, Sports, Technology; (3) Education; (4) Search/Portals/Directories; (5) Services (Careers, Community, Hobbies, ISPs, Mailboxes, Storage); (6) Government; (7) Adult.

The following variables are available for the study:

- Active Time – This is the total time (in seconds) that the user was active in each of the seven categories described above.

---

[16] Our preliminary spider program classified all unique URL domains that have ASCII characters above 192 (this includes all characters with accents marks such as "é" and "û") as French. If such characters were not present, the website was characterized as English. Although, this is a fairly crude mechanism, it probably works reasonably well as a first cut, since most of the URL domains in the data set are in either English or French. Of course, precision demands that we employ a more sophisticated spider program or perhaps a program based on standard language identification schemes. See Grefenstette (1995) for an overview. A more sophisticated spider program will eventually have to characterize all of the approximately one million unique URL full pages.

- Other Time – This is the total active time (in seconds) that the user was active in all other categories[17]

- Age – Age of the user

- Gender – A dummy variable that takes on the value 1 if the user is female and 0 if the user is male.

- Language – A dummy variable that takes on the value 1 if French is the mother tongue of the user and 0 if English is the mother tongue of the user.

- Size – Equal to the number of members of the household, up to a maximum of five. All households with 5 or more members have size equal to five.

- Income – The variable takes on the value 1 if the household income is less than $24,000, 2 if household income is between $25,000 and $40,000, 3 if household income is between $40,000 and $60,000, 4 if household income is between $60,000 and $75,000, 5 if household income is between $75,000 and $100,000, and 6 if household income exceeds $100,000.

- Kids – This is a dummy variable equal to 1 if there are children under age 18 in the household.

- Education – The variable takes on the value 1 if the individual has completed middle school or less, 2 if the individual has attended but not completed high school, 3 if the individual has completed high school, 4 if the individual has attended but not completed college or university, 5 if the individual has an undergraduate degree from a college or university, 6 if the individual has done some post graduate work, but does not have a post-graduate degree, and 7 if the individual has a post-graduate degree.

---

[17] As mentioned above approximately 33 percent of the active time has not yet been categorized. These data are included when computing "Other Time."

- Pereng – This variable is defined to be the percent of the active time that was spent on websites whose content is in English out of the total active time spent at websites whose content is either English or French.[18]

## 3. Methodology

Are there differences in Internet use between native French and native English speakers? We initially focus on Quebec. The reason for doing so is that there may be significant differences among provinces on variables for which we have no control, such as speed of Internet service. Hence, it makes sense to look at Quebec, which is the only province in Canada with significant proportions of native speakers of both English and French in the Media Metrix sample. According to Statistics Canada, there are 602,865 native English speakers in Quebec and 5,728,290 native French speakers in the province. There are 50,585 people who are both native French and English speakers. Approximately 15 percent of our Quebec sample are native English speakers.

The first step is to determine whether native language affects Internet use, where Internet use is defined to be active time spent on the Internet, regardless if the active time is spent on French language or English Language websites. We look at this by category, using the seven categories defined above. The next step involves examining what factors determine the percent of the time that each user spends at English language websites. Initially, we examine this by age group.

## 4. Preliminary Results and Conclusions

Descriptive and summary statistics for the Quebec data are contained in Tables 1-3. Table 1 provides summary statistics for all the variables used in the analysis. The table shows overall, English speakers spent approximately 35 percent more time on the Internet. Table 1 also shows that, on average, native English speaking Quebecois accessed English content websites 87 percent of the time, while native French speaking Quebecois accessed English content websites

---

[18] The spider program could classify not all websites. This is due primarily to the following reasons: (i) the unclassified websites were services that required the user to enter his/her personal ID or information, (ii) the website

64 percent of the time. This already suggests that Quebecois are using the web intensively in English.[19]

Table 2 delineates active time by category. Overall, users spend 36 percent of their active time at search/portal sites, followed by services (21 percent of active time) and retail/business (18 percent of active time). Users spend less than two percent on average at government and educational sites combined.

Table 3 breaks down the active time by age category and shows that there are significant differences among age groups. In the under 15 age group, native French speakers spent approximately 66 percent more time on the Internet than native English speakers, while in the 25-44 age group, native English speakers spent approximately 77 percent more time on the Internet than native French speakers.[20]

In order to examine whether the differences in Active Time vary across categories, Table 4 presents preliminary ordinary least squares (OLS) regression results with Active Time in seconds as the dependent variable. Table 4 shows that native French-speaking Quebecois are somewhat more likely than their English counterparts to use government sites, while English speaking Quebecois are somewhat more likely to spend time at search sites.[21] There is virtually no difference between native French and English speakers in the other categories.[22] These preliminary results suggest that in most categories, native French speakers in Quebec are not less likely than native English speakers to use the Internet.[23]

Table 5 examines what factors determine the percent of the time spent at English language websites. Table 5, which reports ordinary least squares (OLS) regressions, shows that language is much less of a barrier for the youngest users (age less than 15) and for the next youngest group

---

redirected the user more than four times.

[19] There was little difference in this measure among the various age groups.

[20] The age group classification was chosen to match the data from Statistics Canada. See table 6.

[21] The differences are significant at (approximately) the 90 percent level of confidence.

[22] English speakers spent significantly more active time than French speakers in the adult category.

[23] These results are still quite preliminary because 10% of the active time data for Quebec has not yet been categorized.

of users (ages 15 to 24). This is despite the fact that the youngest native French speakers in Quebec are the least likely to have knowledge of English.[24] It goes without saying that these results are also preliminary and further research needs to be undertaken.[25]

While these results are quite preliminary, they suggest that language is much less of a barrier for younger users. These results are consistent with the scenario in which non-native English speakers extensively use English-language sites, the incentives for web sites to make their content available in other languages is reduced, the incentive for individuals to learn English as a second language is enhanced, and that the Internet will promote English as a global language.

The next step will be to incorporate data on the rest of Canada. It will be extremely interesting to compare native English-speaking residents of Quebec to English speaking Canadians in other major provinces, since so many of the native English-speakers in Quebec are bilingual. According to Statistics Canada, Quebec has a much larger portion of bilingual speakers than other major provinces. Are there differences in Internet use patterns between English speakers in Quebec, and other English speakers from other parts of Canada? The same techniques we described in section 3 will be employed to answer this question.

---

[24] See Tables 5 and 6. Of course, our group of under-15 users primarily consists of youths in the 10-15 age group. This group is likely to be much more bilingual than others in the under-15 group in the population.

[25] Given that the dependent variable ranges between 0 and 1, logit regressions will eventually be employed here. The OLS regressions just provide a quick look at the data.

# Appendix: Tables

**Table 1. Descriptive Statistics**

| French speakers, N=819 | | | | |
|---|---|---|---|---|
| VARIABLE | MEAN | STD. DEV | MINIMUM | MAXIMUM |
| Age | 36.02 | 16.15 | 2 [26] | 99 |
| Female | 0.495 | .500 | 0 | 1 |
| Income | 3.27 | 1.51 | 1 | 6 |
| Size | 2.83 | 1.24 | 1 | 5 |
| School | 3.97 | 1.58 | 1 | 7 |
| Kids | 0.44 | 0.50 | 0 | 1 |
| Active time | 34,154.43 | 53,852.03 | 1 | 849,625 |
| Pereng | 0.87 | 0.17 | 0 | 1 |
| English Speakers N=163 | | | | |
| VARIABLE | MEAN | STD. DEV | MINIMUM | MAXIMUM |
| Age | 35.91 | 15.56 | 7 | 74 |
| Female | 0.54 | 0.50 | 0 | 1 |
| Income | 3.36 | 1.71 | 1 | 6 |
| Size | 3.01 | 1.41 | 1 | 5 |
| School | 4.11 | 1.62 | 1 | 7 |
| Kids | 0.38 | 0.49 | 0 | 1 |
| Active Time | 46,133.61 | 66,164.52 | 9 | 402,622 |
| Pereng | 0.64 | 0.25 | 0 | 1 |

---

[26] Three French speakers are listed as "99" years old. No one else is over 79 years old. Similarly there is a small group of French speakers under the age of 5. Nothing qualitative changes if we ignore these two groups in the analysis.

## Table 2. Percent of Total Active Time by Category

|  | Percent of Total Active Time |
|---|---|
| Retail, Business | 18.9 |
| Entertainment | 10.6 |
| Education | 0.9 |
| Search | 35.7 |
| Services | 21.1 |
| Government | 1.9 |
| Adult | 10.9 |

## Table 3: Active Time by Age, Language, and Category (E=English, F=French).

| Age group | | Mean Active Time By Category | | | | | | | |
|---|---|---|---|---|---|---|---|---|---|
| | | **Overall** | Retail, Business | Entertain | Edu | Search | Services | Govt | Adult |
| Under 15 | E | **9,564** | 890 | 2162 | 104 | 5848 | 1876 | 192 | 676 |
| | F | **15,900** | 2119 | 3804 | 308 | 5215 | 4389 | 340 | 1168 |
| 15-24 | E | **39,741** | 3890 | 5052 | 1628 | 14010 | 6533 | 485 | 4792 |
| | F | **27,707** | 3877 | 3810 | 1029 | 9176 | 5343 | 1691 | 4130 |
| 25-44 | E | **64,632** | 14520 | 9094 | 851 | 17072 | 11678 | 988 | 7586 |
| | F | **36,483** | 7880 | 4672 | 732 | 10501 | 8408 | 1309 | 4035 |
| 45-64 | E | **54,976** | 10524 | 4671 | 698 | 16937 | 7257 | 1381 | 11422 |
| | F | **48,707** | 9004 | 6681 | 712 | 14659 | 8575 | 1423 | 4347 |

# Table 3: Preliminary Regression Results

The Dependent Variable is Active Time in the Category. The t-values are in parentheses. (Bold fonts mean significance at the .05 level.)

|  | Retail, Business | Entertain | Edu | Search | Services | Govt | Adult |
|---|---|---|---|---|---|---|---|
| Constant | -2073.94 (-0.73) | 2793.33 (1.15) | 71.33 (0.11) | 11482.38 (3.46) | -4623.1 (-1.30) | 1157.5 (1.87) | 10509.7 (3.84) |
| Age | 21.26 (0.59) | 35.36 (1.17) | -6.02 (-0.76) | 67.39 (1.64) | 54.80 (1.23) | 3.01 (0.39) | -35.58 (-1.00) |
| Female | 214.39 (0.21) | 705.22 (0.81) | -43.21 (-0.19) | 425.14 (0.36) | **2528.63** **(1.97)** | -136.16 (-0.62) | **-5593.7** **(-5.66)** |
| Language | -912.37 (-0.68) | -510.64 (-0.45) | -118.24 (-0.41) | -2441.23 (-1.56) | 1810.15 (1.06) | 494.67 (1.62) | **-3303.7** **(-2.51)** |
| Size | **-1574.23** **(-2.66)** | -278.50 (-0.55) | 142.65 (1.19) | -392.10 (-0.57) | -163.88 (-0.22) | -48.99 (-0.40) | -181.09 (-0.32) |
| Income | **1139.70** **(3.26)** | 236.23 (0.77) | 2.81 (0.04) | **-941.40** **(-2.31)** | **-1314.1** **(-2.99)** | **-204.18** **(-2.75)** | -148.57 (-0.45) |
| School | **729.20** **(2.11)** | -436.01 (-1.46) | **216.23** **(2.68)** | -513.71 (-1.29) | 654.03 (1.51) | 50.91 (0.69) | 14269 (0.41) |
| Kids | 2180.96 (1.48) | 860.86 (0.68) | **-600.55** **(-2.01)** | -263.67 (-0.15) | **5758.6** **(3.12)** | -60.81 (0.20) | -1572.1 (-1.11) |
| Other time | **0.18** **(18.05)** | **0.066** **(8.80)** | -.00005 (-0.03) | **0.21** **(16.65)** | **0.21** **(15.98)** | **.0052** **(3.44)** | **.027** **(3.83)** |
| # of Obs | 888 | 820 | 330 | 934 | 913 | 505 | 472 |
| Adj $R^2$ | 0.29 | 0.09 | .01 | 0.25 | 0.24 | 0.04 | 0.10 |

# Table 4: Preliminary Regression Results

Dependent Variable is Pereng. These regressions are done at the level of the individual by agegroup. The t-values are in parentheses. (Bold fonts mean significance at the .05 level.)

|  | Age Less than 15 | Ages 15 to 24 | Ages 25 to 44 | Ages 45 to 64 |
|---|---|---|---|---|
| Constant | 0.53 (2.87) | 1.24 (6.46) | 0.94 (9.28) | 0.73 (3.88) |
| Priage | **0.022 (2.49)** | **-0.019 (-1.94)** | -0.0037 (-1.67) | 0.0028 (0.89) |
| Active Time | 3.78e-07 (0.32) | 2.26e-07 (0.48) | 4.88e-08 (0.25) | 1.79e-07 (0.75) |
| Female | -0.045 (-0.75) | -0.032 (-0.85) | -0.0074 (-0.30) | -0.043 (-1.49) |
| Language | -0.14 (-1.49) | **-0.18 (-3.69)** | **-0.22 (-6.70)** | **-0.24 (-6.31)** |
| Size | 0.025 (0.56) | -0.0044 (-0.22) | -0.0075 (-0.48) | -0.012 (-0.68) |
| Income | -0.019 (0.87) | -0.0020 (-0.15) | 0.012 (1.33) | 0.012 (1.06) |
| School | 0.034 (0.66) | 0.0097 (0.45) | 0.0028 (0.31) | -0.00042 (-0.04) |
| Kids | 0.022 (0.14) | 0.0025 (0.05) | 0.017 (0.44) | 0.045 (0.90) |
| # of Obs. | 100 | 162 | 418 | 269 |
| Adj $R^2$ | 0.08 | 0.09 | 0.09 | 0.13 |

## Table 5: Mother Tongue by Region

Source: Statistics Canada (1996)

| Region  | English    | French    | Eng & French | % English |
|---------|-----------:|----------:|-------------:|----------:|
| BC      | 2,827,325  | 54,020    | 6,655        | 98        |
| Ontario | 7,825,360  | 484,620   | 33,940       | 94        |
| Quebec  | 602,865    | 5,728,290 | 50,585       | 10        |
| Other   | 5,884,610  | 405,570   | 17,925       | 94        |
| Total   | 17,140,160 | 6,672,500 | 117,165      | 72        |

## Table 6: Percent of Quebec Population with Knowledge of Other Official Language (By Age Group)

Source Statistics Canada (1996).

| Mother Tongue | Overall | Less than 15 | 15-24 | 25-44 | 45-64 |
|---------------|--------:|-------------:|------:|------:|------:|
| English       | 40      | 38           | 66    | 48    | 35    |
| French        | 32      | 4            | 39    | 41    | 36    |